\documentclass[aps,prl,twocolumn,showpacs,superscriptaddress,groupedaddress]{revtex4}  
\usepackage{graphicx}  
\usepackage{dcolumn}   
\usepackage{bm}        
\usepackage{amssymb}   
\usepackage{epstopdf}
\usepackage{color}
\usepackage{dcolumn}
\usepackage{bm}
\def\gapx{\lower 2pt \hbox{$\buildrel>\over{\scriptstyle{\sim}}$\ }}
\def\lapx{\lower 2pt \hbox{$\buildrel<\over{\scriptstyle{\sim}}$\ }}

\RequirePackage[colorlinks,linkcolor=black,citecolor=blue,hyperindex]{hyperref}

\begin{document}
\title{Superglass phase of interaction-blockaded gases on a triangular lattice}

\author{Adriano Angelone} 
\affiliation{icFRC, IPCMS (UMR 7504) and ISIS (UMR 7006), Universit\'e de Strasbourg and CNRS, 67000 Strasbourg, France}%
\author{Fabio Mezzacapo}%
\affiliation{icFRC, IPCMS (UMR 7504) and ISIS (UMR 7006), Universit\'e de Strasbourg and CNRS, 67000 Strasbourg, France}%
\author{Guido Pupillo}%
\affiliation{icFRC, IPCMS (UMR 7504) and ISIS (UMR 7006), Universit\'e de Strasbourg and CNRS, 67000 Strasbourg, France}%

\date{\today}

\begin{abstract}
We investigate the quantum phases of monodispersed bosonic gases confined to a triangular lattice and interacting via a class of soft-shoulder potentials. The latter correspond to soft-core potentials with an additional hard-core onsite interaction. Using exact quantum Monte Carlo simulations, we show that the low temperature phases for weak and strong interactions following a temperature quench are a homogeneous superfluid and a glass, respectively. The latter is an insulating phase characterized by inhomogeneity in the density distribution and structural disorder. Remarkably, we find that for intermediate interaction strengths a {\it superglass} occurs in an extended  region of the phase diagram, where glassy behavior coexists with a sizable finite superfluid fraction. This glass phase is obtained in the absence of geometrical frustration or external disorder and is a result of the competition of quantum fluctuations and cluster formation in the corresponding classical ground state. For high enough temperature, the glass and superglass turn into a floating stripe solid and a supersolid, respectively. Given the simplicity and generality of the model, these phases should be directly relevant for state-of-the-art experiments with Rydberg-dressed atoms in optical lattices.  
\end{abstract}

\pacs{32.80.Ee, 67.80.K-, 05.30.Jp, 61.43.Fs}

\maketitle

It is well established that bosonic and fermionic systems subjected to a disordered external potential feature localization phenomena~\cite{Anderson,Review}. The interplay between disorder, interactions and many-body quantum effects such as superfluidity  is now a subject of intense research \cite{manybodyloc1,manybodyloc2,manybodyloc3,manybodyloc4, nussinov0, nussinov}, as, e.g., bosons in random environments occur in a variety of experimentally relevant systems ranging from cold atoms  \cite{Damski2003,Billy2008, Sanchez2008, Roati2008, White2009}, to superconductors \cite{Gold1998} and quantum liquids \cite{Balibar2008,Reppy2007}.  Usually, the combination of disorder and repulsive interactions inhibits the emergence of superfluidity and Bose-Einstein condensation (BEC) and leads to an insulating gapless phase, known as Bose glass \cite{Fisher1989, Krauth1991, Pollet2009}.

Remarkably, results of quenched Monte Carlo simulations in the context of $^4$He have shown that superfluidity and BEC may coexist with structural disorder and inhomogeneity (i.e., glassy physics)  in the absence of any random external potentials \cite{Boninsegni2006}. The resulting out-of-equilibrium state was termed {\it superglass} (SG), as a disordered analog of the supersolid (SS) phase \cite{BoninsegniRMP}. While experiments have so far remained inconclusive \cite{Hunt2009,West2009}, this proposal has spurred considerable theoretical activity to derive possible microscopic models of a SG \cite{Biroli2008,Carleo2009,Dang2009,Melko2010,Biroli2011,Mueller2011,Larson2012}. Exact numerical results for bosons on lattices have shown that a thermodynamic SG phase can indeed emerge as a result of a competition of quantum fluctuations and externally induced frustration. For attractive interactions the latter can be induced via a random chemical potential \cite{Dang2009}, while for repulsive ones a SG can occur in theoretical models where either a self-disordered environment is induced by geometrical frustration (e.g., on random graphs)  \cite{Carleo2009} or where disorder is a consequence of properly chosen random inter-particle interactions \cite{Melko2010,Larson2012}. In this context, main open questions are whether it is possible to obtain a SG in any theoretical models where frustration is not artificially built in the Hamiltonian, and if this new phase of matter may be experimentally observable in any physical system.

Here, we show that the SG phase can exist for a large class of bosonic lattice Hamiltonians. The latter are of the extended Hubbard-type, featuring a soft-shoulder interaction potential. Surprisingly, glassy behavior is obtained  in the absence of any externally imposed frustration e.g., in the lattice geometry, or in the interactions. Rather, frustration is here induced by cluster formation for large particle density, similar to the conditions of SS formation in soft-core models \cite{Rica1994,Henkel2010,Cinti2010}. As an example, we consider a simple triangular lattice  with isotropic two-body interactions. We analyze  the  phases   and, following a quench in the temperature $T$ or in the interaction strength,  demonstrate the existence of both a classical glass (G) and a SG at low enough  $T$. The latter are the out-of-equilibrium counterparts of a floating stripe solid (S) and a SS, respectively. These glass and superglass phases should be observable in experiments with Rydberg-dressed  alkali atoms loaded into optical lattices. 

The relevant Hamiltonian for hard-core bosons confined to a 2D triangular lattice reads
\begin {equation}\label{eq:Ham}
\mathcal{H} = - t \sum_{\{ i,j \}} \left( {b}^{\dagger}_i {b}_j + {b}^{\dagger}_j {b}_i \right) + V \!\!\!\!\! \sum_{i<j; \, \,{r}_{ij} \leq r_c} \! \!\!{n}_i {n}_j.
 \end{equation}
Here, $b_i$ ($b^\dagger_i$) are hard-core bosonic annihilation (creation) operators at site $i$, ${n}_i= {b}^{\dagger}_i {b}_i$, $r_{ij}$ is the distance between sites $i$ and $j$, and $t$ is the tunneling rate on a lattice of spacing $a$. In the following, $t$ and $a$ are used as units of energy and length, respectively. In classical physics, the {\it soft-shoulder} potential of Eq.~(\ref{eq:Ham}) is of interest for soft-matter models of, e.g., colloids \cite{Mladek2006,Lenz2012, Sciortino2013}. In quantum physics, this potential can be engineered in clouds of cold Rydberg atoms, where both the strength $V$ and the range  $r_c$ of the interaction can be tuned by weakly-admixing the Rydberg level to the ground state~\cite{Santos,Honer, Pupillo2010,Henkel2010,Cinti2010,Henkel2012,Macri2014,Biedermann2015, Helmrich2015} (see SupMat). The additional onsite hard-core constraint can be enforced using, e.g., Feshbach resonances. 

The quantum phases of Eq.~(\ref{eq:Ham}) with $r_c=1$ (i.e., nearest-neighbor interactions) are well known \cite{Murthy97, Boninsegni2003, Melko2005, Wessel2005, Boninsegni2005}: for densities $\rho < 1/3$ ($\rho > 2/3$), $\rho = 1/3$ ($\rho = 2/3$) and $\rho > 1/3$ ($\rho < 2/3$) the low-energy phase is a superfluid (SF), a gapped lattice S, or a gapless SS, respectively. The latter is an exotic state of matter where density correlations (here with $\sqrt{3} \times \sqrt{3}$ ordering) coexist with a finite superfluid fraction $\rho_s$, which is a result of doping the solid with interstitials (vacancies). The SS phase is generally robust against perturbations to the Hamiltonian~(\ref{eq:Ham}), and may be observed experimentally, e.g., with cold quantum gases trapped in optical lattices and interacting via dipolar interactions \cite{Pollet2010,Capo2010, WL2015}.

In this work, we are interested in Eq.~(\ref{eq:Ham}) with $r_c>1$. For  $r_c>1$ the interactions belong to a large class of potentials that support the formation of self-assembled clusters of particles for sufficiently large densities $r_c\sqrt{\rho} >1$ \cite{Lenz2012, Sciortino2013}. Such a phenomenon is essentially independent of the details of the interactions, as long as the latter display a negative Fourier component \cite{Mladek2006}.
In the classical regime (i.e., $t=0$)   cluster formation has been shown to lead to frustration, which is manifested in an exponential growth of the ground state degeneracy as a function of the system size \cite{Mattioli2013}.
 In the quantum regime (i.e., $t>0$) this leads to several novel exotic phenomena at equilibrium: anomalous Luttinger-Liquid behavior \cite{Mattioli2013} and emergent supersymmetry in 1D lattice geometry \cite{Dalmonte2015} as well as free-space supersolidity in 2D \cite{Cinti2014,Henkel2010,Cinti2010}. The latter occurs, for appropriate values of interaction strength, at any density fulfilling the clusterization condition $r_c \sqrt{\rho}>1$ \cite{Cinti2014}. In the following we consider, as a way of example, the simplest cluster forming potential with $r_c=2$ and  incommensurate particle densities consistent with such a condition. Our main focus is the demonstration of a G and SG emerging when a crystal and a SS are driven out of equilibrium via a temperature quench, respectively. Glassy phases for different $\rho$, $r_c$ and quench protocols are discussed in the Supplementary Material (SM).
 
We study the Hamiltonian in Eq.~(\ref{eq:Ham}) by means of Path Integral Quantum Monte Carlo simulations based on the worm-algorithm [\onlinecite{Prok98}]. This technique is numerically exact for bosonic systems and allows for accurate estimates of the superfluid fraction $\rho_s  =  \langle ( W_x^2 + W_y^2)/(6 \beta \rho)\rangle$ and the static structure factor $S(\mathbf{k})/N = \sum_{i,j} \exp [-i \mathbf{k} \cdot (\mathbf{r}_i - \mathbf{r}_j)] \langle {n}_i {n}_j \rangle /N^2$ ($N$ is the lattice size). The latter measure superfluidity and diagonal crystalline orders, respectively. Here, $\beta=1/(k_BT)$, with $k_B$ Boltzmann constant (in the following set to 1), $W_i$ is the winding number in the $i$-th direction,  $\mathbf{k}$ is a lattice wave-vector, and $\langle\cdots\rangle$ stands for statistical average. In addition, we compute the renormalized Edwards-Anderson order parameter $\widetilde{q}_{EA} = q_{EA}/q_{EA}^{\rm max}$, which, in the absence of crystalline order, is the well-accepted observable to identify glassy behavior on a lattice \cite{Carleo2009, Binder86}. Here, $q_{EA}=\sum_{i = 1}^N \langle n_i - \rho \rangle^2$ and $q_{EA}^{\max}=N \rho ( 1 - \rho)$ is its  maximum value obtained for a classical situation with no particle delocalization. We perform large-scale simulations with up to $N=2304$ lattice sites and temperatures as low as $T/t=1/12$. For each $N$ and $T$, numerical values for the observables above are obtained by averaging over a minimum of 32 and a maximum of 100 different realizations of the quench.

\begin{figure}[t]
\centerline{\includegraphics[width = 0.85\columnwidth]{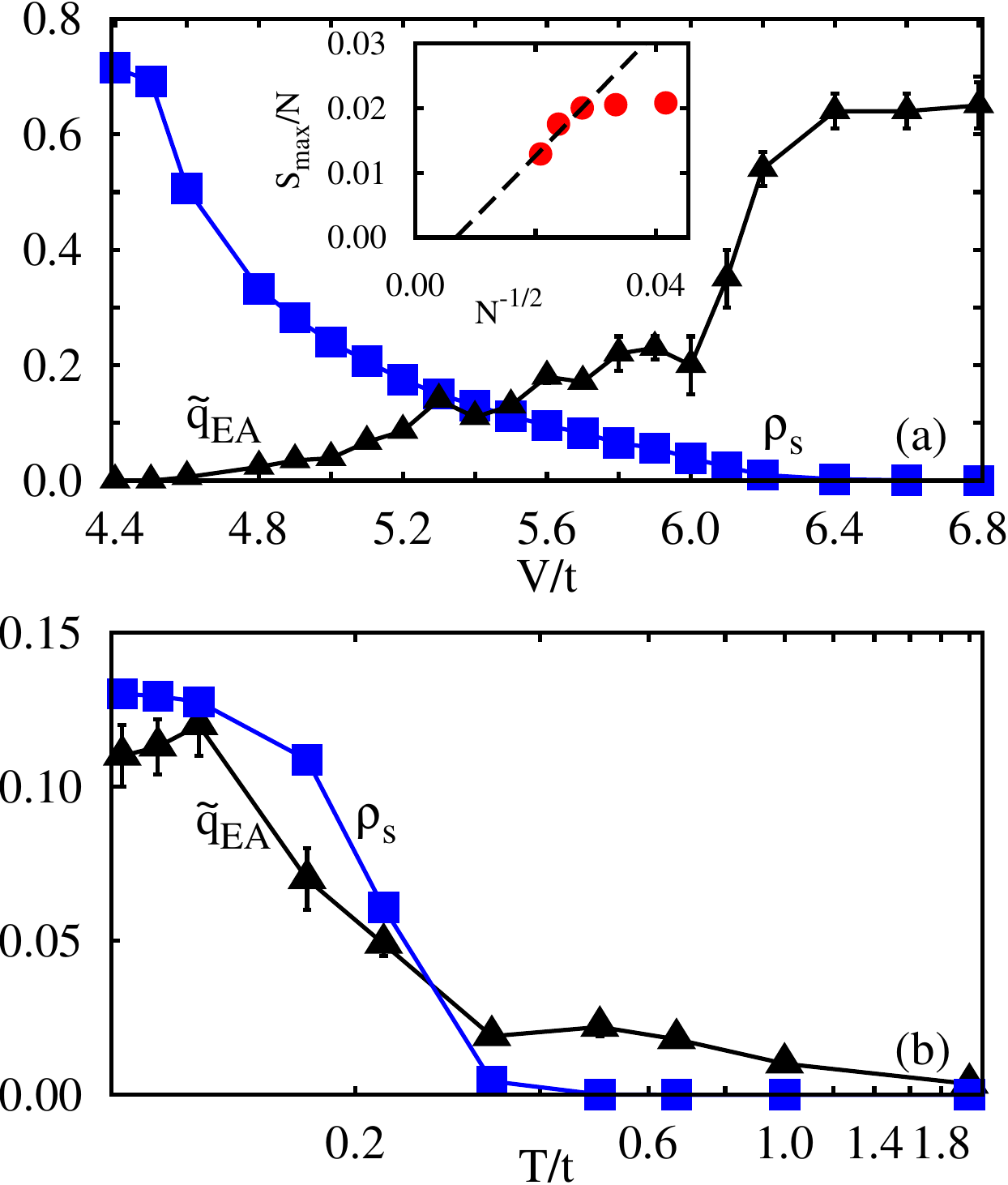}}
\caption{(color online). (a): Superfluid fraction $\rho_s$, and renormalized Edwards-Anderson parameter $\widetilde{q}_{EA}$ as a function of $V/t$, for $T/t=1/12$. (b): $\rho_s$ and  $\widetilde{q}_{EA}$ as a function of $T/t$, for $V/t=5.4$. In both panels the density is $\rho=13/36$ and the lattice size is $N=900$. Solid lines are guides to the eye. Inset: maximum value of the structure factor $S_{\rm max}/N$ as a function of $1/\sqrt{N}$ for $\rho=13/36$, $V/t=5.4$ and $T/t=1/12$; the dashed line is a linear fit for the three largest system sizes.} 
\label{Fig1}
\end{figure}

\begin{figure}[t]
\centerline{\includegraphics[width = 0.85\columnwidth]{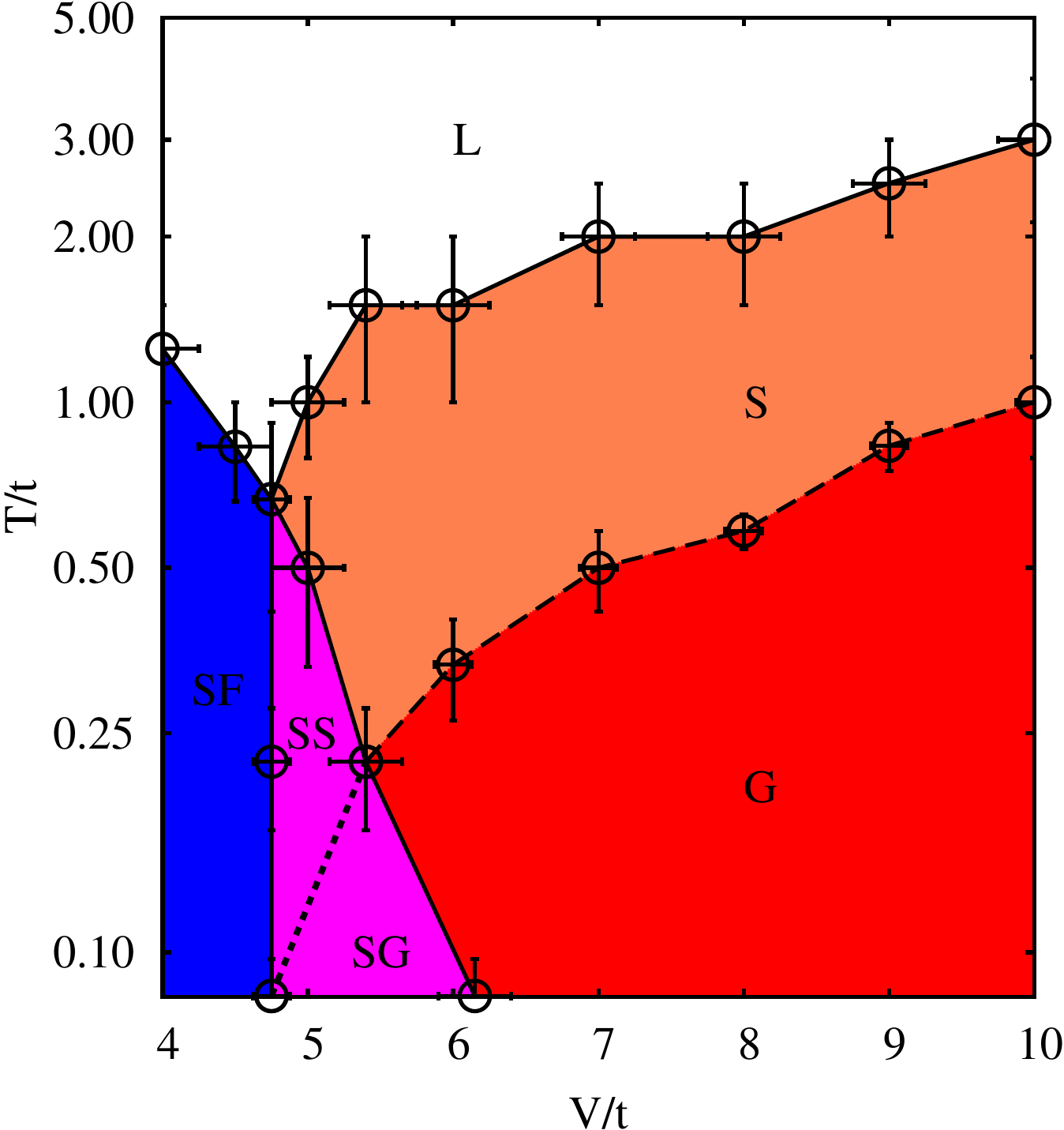}}
\caption{(color online) Phase diagram of model Eq.~(1)  with $r_c=2$ as a function of temperature $T/t$ and interaction strength $V/t$, for  particle density $\rho=13/36$. Equilibrium phases: normal liquid (L), superfluid (SF), stripe-crystal (S), and supersolid (SS). A temperature quench to final values of $T/t$ below the dashed line leads to a glass (G). The existence of a superglass (SG) is demonstrated below the dotted line using the same quench protocol.} 
\label{fig:PD}
\end{figure}
Figure~\ref{Fig1} [panel (a)] shows example  results for the superfluid fraction $\rho_s$ and the renormalized Edwards-Anderson parameter $\widetilde{q}_{EA}$  as a function of the interaction strength $V/t$  for $N=900$ and $T/t=1/12$.  Within the interesting range  of interaction ($5.0 \lesssim V/t \lesssim 6.0$),   $\rho_s$ is found to decrease monotonically with increasing  $V/t$ from approximately  $0.25$ to about $0.05$.  In the same parameter range, $\widetilde{q}_{EA}$  increases up to values of the order of $\sim 0.2$. We note that in this regime the  system does not feature crystalline order, i.e., the computed structure factor $S(\mathbf{k})/N$ vanishes for any non trivial wave vector $\mathbf{k}\neq 0$ in the thermodynamic limit, as proven in the Inset, where we show the scaling with $N^{-1/2}$ of $S_{\rm max}/N$, the average of the largest peaks of the structure factor over several quench realizations. In the same limit the superfluid fraction stays finite. These data demonstrate one of the main results of this work, namely the  existence, in an extended region of parameters, of a SG, corresponding to an {\it inhomogeneous non-crystalline superfluid}.  The dependence on $T/t$ of both $\widetilde{q}_{EA}$ and $\rho_s$  is shown in Fig.~\ref{Fig1} [panel (b)]. For the specific value of the interaction strength in the panel superglassiness is realised below $T/t \simeq 0.2$.

For weak interactions, the SG phase {\it quantum melts} into a regular homogeneous superfluid (SF) with $\rho_s>0$ and $\widetilde{q}_{EA} \simeq S({\bf k})=0$. For the parameters of Fig.~\ref{Fig1}(a) this is obtained by decreasing the interaction strength below $V/t\simeq 4.8$. On the other hand, sufficiently large interaction strengths are found to inhibit superfluidity and turn the SG into an insulating G. The latter is characterized by a finite value of $\widetilde{q}_{EA}$ and $\rho_s\simeq S({\bf k})= 0$ [i.e., $V/t \gtrsim 6.2$ in the figure]. Within this glass phase quantum effects are largely suppressed. While classical glasses are well known to appear in disordered spin models, as well as in certain polydispersed systems of particles \cite{Binder86}, here we demonstrate that (classical) glassy physics may emerge in a simple and rather general model  of immediate experimental interest for bosons on a regular lattice. 

\begin{figure}[t]
\centerline{\includegraphics[width = 0.85\columnwidth]{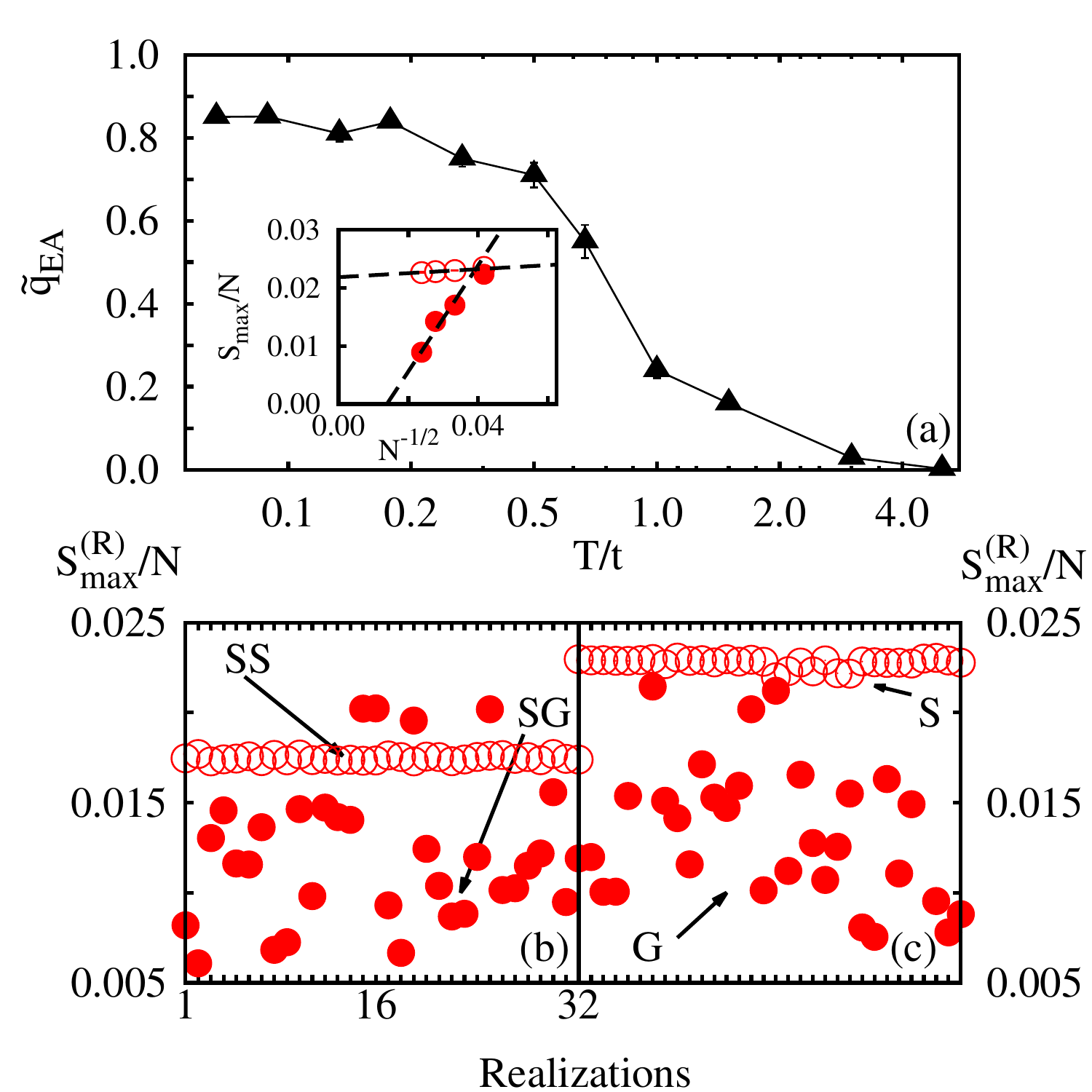}}
\caption{(color online) (a): Normalized Edwards-Anderson order parameter $\widetilde{q}_{EA}$ for model Eq.~(\ref{eq:Ham}) as a function of temperature $T$. The value of the interaction strength is $V/t=10$. Data are for a lattice with size $N=900$ and density $\rho=13/36$. The solid line is a guide to the eye. Inset: Size dependence of $S_{\max}/N$ for $V/t$=10. Values of the temperature are $T/t=1/12$ and $T/t=3/2$ for the full and empty symbols, respectively. Dashed lines are fits to our numerical data. (b-c): Maximum value of the structure factor $S^{(R)}_{\max}/N$ obtained in a given realization of a quench, plotted as a function of the number of different quench realizations. The latter only differ in the (random) initial condition and in the thermalization seed. The corresponding phases in the thermodynamic limit are indicated. Fluctuations in the values of $S^{(R)}_{\max}/N$ indicate glassy behaviour. The parameters are: $N = 1764$, $T/t = 1/2.5$, $V/t = 5$ (SS), $N = 2304$, $T/t = 1/9$, $V/t = 5.4$ (SG), $N = 1296$, $T/t = 1/0.7$, $V/t = 10$ (S), $N = 1296$, $T/t = 1/12$, $V/t = 10$ (G).}
\label{Fig3}
\end{figure}

The computed  phase diagram of  Eq.~(\ref{eq:Ham}) is shown in Fig.~\ref{fig:PD} for a choice of particle density $\rho=13/36$ as a function of $T/t$ and $V/t$. At high temperatures  we find a normal liquid (L) phase  independently of the values of $V/t$, as expected. For sufficiently small interaction strength $V/t\lesssim 4.8$, this normal phase turns into a homogeneous superfluid  by decreasing $T/t$, via a phase transition which is consistent with the Berezinskii-Kosterlitz-Thouless scenario. On the other hand, for large enough $V/t$ and following a quench to low $T$ the system displays a marked insulating glassy behavior with $\widetilde{q}_{EA} \neq 0$, $S({\bf k})=\rho_s =0$ [Fig.~\ref{Fig3}(a) and full symbols in the Inset]. The interplay between glassy physics and superfluidity is mostly evident for values of $T/t$ below the dotted lines in Fig.~\ref{fig:PD}, resulting in the SG scenario discussed above.

Interestingly, we find that fluctuations can restore crystalline order for sufficiently large $T$. This is shown for intermediate temperatures in Fig.~\ref{fig:PD}, where a S (SS) phase intervenes between the low-temperature G (SG) and the high-temperature normal L. Here, the crystal is a floating stripe solid, with finite diagonal long range order in the thermodynamic limit. Examples for the finite size scaling of the maximum value of the structure factor $S_{\rm max}/N$ in the S and G phases are shown in the inset of  Fig.~\ref{Fig3}(a) (empty and full symbols, respectively). While in the S phase $S_{\rm max}/N$ is essentially independent of the system sizes investigated in this work, in the G phase $S_{\rm max}/N$  vanishes in the thermodynamic limit. In both phases  $\rho_s \simeq 0$.

The  difference between the glassy and crystalline phases is shown in Fig.~\ref{Fig3}(c), where we plot the maximum value of the structure factor  $S^{(R)}_{\rm max}/N$ for each individual realization of a temperature quench at a given $V/t$ and final $T/t$. In the crystalline phases,  $S^{(R)}_{\rm max}/N$ is essentially identical in all realizations and $S_{\rm max}/N$ remains finite in the thermodynamic limit. However, within the glassy phases, $S^{(R)}_{\rm max}/N$ can fluctuate widely and in average decreases to zero with the system size. As shown in Fig.~\ref{Fig3}(b) the dependence of $S^{(R)}_{\rm max}/N$ on the realization for the SG and the SS phase is similar to that for the G and the S ones, respectively.

\begin{figure}[t]
\centerline{\includegraphics[width=0.8\columnwidth]{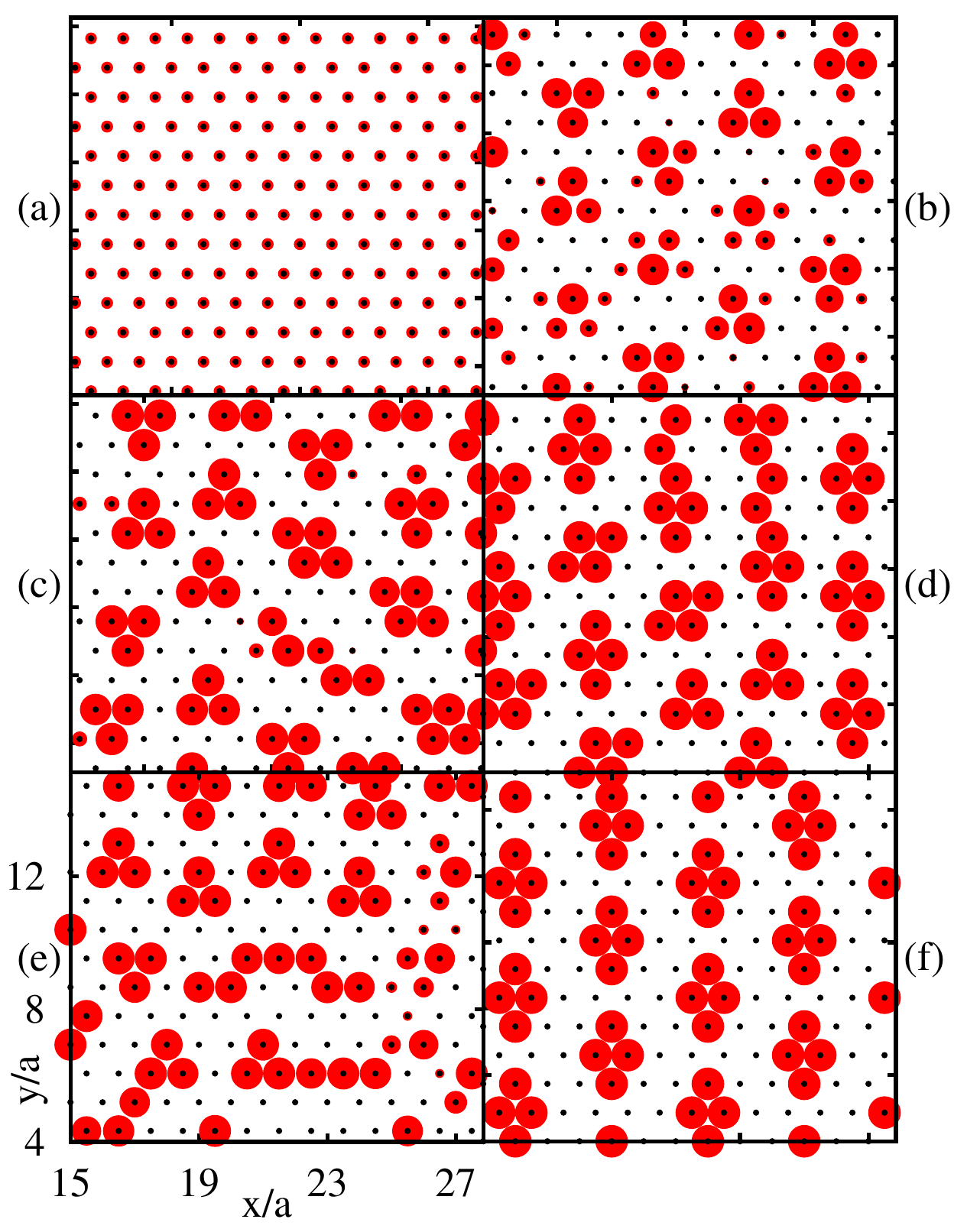}}
\caption{(color online) Averaged site density for a portion of the system. Black circles depict the lattice sites. Density values are proportional to the size of red circles. Panel (a) shows an homogeneous superfluid phase ($T/t=1/9$ and $V/t=4$); panels (b) and (c),  refer to a SG ($T/t=1/12$, $V/t=5.4$), and to a normal G ($T/t=1/12$, $V/t=10$), respectively. In panels (a-c) the density is $\rho=13/36$. Panels (d) and (e) show the glassy density map obtained for $\rho=0.401$  ($V/t=10$, $T/t=1/12$), and for the same density of panels (a-c),   using the van der Waals purely repulsive soft potential  with $V_{ryd}/t=30$, $T/t=1/3$ (see details in SM), respectively. Panel (f): a crystalline structure stabilizable at $\rho=1/3$, $V/t=10$ and $T/t=1$. } 
\label{Fig4}
\end{figure}

Further insight into the phases of Hamiltonian (1) is given by the analysis of the averaged site-density maps  in Fig.~\ref{Fig4}. 
Specifically, we  show results for a portion of the system and for a choice of $T/t$ and $V/t$ such that the system is a SF [panel (a)], a SG [panel (b)], and a G [panels (c-e)]. For comparison, panel (f) shows a cluster-type crystalline phase [i.e., $S(\mathbf{k}) \neq 0$]  stabilizable at a density $\rho=1/3$, for $V/t=10$ and $T/t=1$~\cite{Note100}.  
In the homogeneous SF the average occupation number at each site equals the density $\rho$ of the system, as expected. The resulting value of $\widetilde{q}_{EA}$ is thus negligible. Conversely, when $V/t$ is large [panel (c)] the spatial density is highly inhomogeneous: particles form self-assembled clusters characterized by different numbers of constituents and spatial orientations, as well as by varying inter-cluster distances. These features lead, in the thermodynamic limit, to the absence of diagonal long range order, similar to an (emergent) polydispersity. 
Noticeably, the occupation number of lattice sites between clusters is here substantially suppressed, signaling particle localization. 
The resulting glass phase is insulating, similarly to, e.g, a regular Bose glass obtained by externally induced disorder~\cite{Fisher1989}. 

Figure~\ref{Fig4}(b) shows that cluster formation (i.e., inhomogeneity) persists even at intermediate values of $V/t$, leading  to a non zero value of $\widetilde{q}_{EA}$ in the absence of crystalline order. The occupation of inter-cluster lattice sites is here enhanced with respect to panel (c). Such an enhancement is  due to the presence of quantum fluctuations and exchanges of identical particles, responsible for the finite value of $\rho_s$ and thus of superglassiness.

While here we have mainly focused on the minimal model Hamiltonian with $r_c=2$ and a given density $\rho=13/36$,  we have verified that glassy phases also occur for larger $r_c$ and densities $\rho$ satisfying the clusterization condition,  as well as when the details of the cluster forming potential are changed (see Fig.~\ref{Fig4}  and SM). In particular, glassy phases are obtained for soft-core van der Waals type interactions  relevant to experiments with Rydberg dressed atoms~\cite{Biedermann2015,Gross} [see Fig.~\ref{Fig4} and SM]. In view of possible experiments, where $V/t$ can be easily varied, we have verified that [see SM] (i) quenches in $V/t$ for a fixed low $T$ from a superfluid lead to the same SG and G phases described above; in addition, (ii) the SG and G phases are robust against density variations at the percent level. In particular, the G phase is found at all densities explored, up to $\rho\simeq 0.4$, for sufficiently large $V/t$.  We emphasize that while the finite lifetime of a Rydberg-dressed gas can be a limitation to the observability of phases in thermal equilibrium, glassy phases are realized out of equilibrium and their relevant properties are essentially those emerging in the first stages of the experiment.

The glass and superglass  phases discussed here are the  low-$T$ quenched counterparts of the equilibrium solid and supersolid found for intermediate $T$. These latter  equilibrium phases are analogous to those found in, e.g., Ref.~\cite{Cinti2014} in continuous space. While recent results point to the existence of out-of-equilibrium glassy-type phenomena  in the classical regime for cluster-forming systems ~\cite{Diaz-Mendez2014} (see also Ref.~\cite{Diaz-Mendez2015} for a discussion on the dynamics of the equilibrium phases), the existence of glassy dynamics in the quantum regime and in continuous space remains an open question. 

In conclusion, we have demonstrated that glassy phases can be realized for a broad class of simple bosonic frustration-free Hamiltonians of the extended  Hubbard-type. For intermediate interaction strength  the interplay between  quantum fluctuations, statistics and glassy physics gives rise  to an exotic SG scenario, where glassiness coexists with superfluidity, in contrast to a conventional Bose glass. In our model, frustration arises from the self-assembling of clusters, which is a direct consequence of the (isotropic) inter-particle interaction potential at high enough density. The physics described in this work should be directly relevant for experiments with ultracold Rydberg-dressed atoms confined to optical lattices \cite{Henkel2010, Macri2014,Lesanovsky2013}. We hope that our work will provide new insights for unveiling other general mechanisms  to glassy physics, and in general to frustration-induced phenomena  both in the classical and quantum regime. Interesting extensions might include the search for exotic phenomena beyond the SG, such as frustration-induced Bose metals~\cite{Fisher2011,Rigol2011} and emergent gauge fields~\cite{MilaBook}.\\

We thank  B. Capogrosso-Sansone, G. Carleo, R. Diaz-Mendez, N. V. Prokof'ev, and F. Zamponi for discussions. Research is supported by the European Commission via ERC-St Grant ‘‘ColdSIM’’
(No. 307688). We acknoledge additional partial support from EOARD,  H2020 FET Proactive project RySQ (grant N. 640378),  ANR-FWF via ``BLUSHIELD", and UdS via Labex NIE and IdEX, computing time at the HPC-UdS.

\newpage
\newpage
\widetext
\setcounter{equation}{0}
\makeatletter 
\renewcommand{\theequation}{S\@arabic\c@equation}
\makeatother

\setcounter{figure}{0}
\makeatletter 
\renewcommand{\thefigure}{S\@arabic\c@figure}
\renewcommand{\bibnumfmt}[1]{[S#1]}
\renewcommand{\citenumfont}[1]{S#1}
\makeatother

\onecolumngrid
\begin{center}

{\bf \large Supplementary Material to\\ ``Superglass phase of interaction-blockaded gases on a triangular lattice"}\\
\vspace{0.5cm}
{Adriano Angelone$^1$, Fabio Mezzacapo$^1$, and Guido Pupillo$^1$}\\
{$^1$icFRC, IPCMS (UMR 7504) and ISIS (UMR 7006),\\ Universit\'e de Strasbourg and CNRS, Strasbourg, France}
\end{center}

{\small We demonstrate the occurrence of glassy physics  in the model Hamiltonian introduced in  Eq.~(1) of the main text for different choices of interaction radius $r_c > 1$  and  particle density $\rho$ fulfilling the clusterization condition $r_c\sqrt{\rho} > 1$. We find no evidence of a glass for $r_c=1$, i.e., when cluster formation does not occur. These results indicate that glassiness is favored by the formation of self-assembled clusters. Low-temperature glassy scenarios are found following either a quench in temperature $T$ or in the interaction strength $V/t$. Glassy phases are robust against variations of the details of the cluster-forming interaction potential. Parameters for a possible realization of glassy dynamics with Rydberg-dressed gases trapped in optical lattices are discussed.}

\null\vspace{0.5cm}\null
\twocolumngrid

 \subsection{Dependence of glassy behavior on the interaction radius $r_c$, density $\rho$ and  cluster formation}
In this section we investigate the dependence of the glass behaviour found in the main text on the  radius $r_c$ of the interaction  potential of Hamiltonian (1) and on the particle density $\rho$. In particular, we are interested in exploring whether  the appearance of glassiness  is related to the formation of clusters in the ground state. 

By performing extensive quantum Monte-Carlo calculations, we demonstrate that glassy phases also occur for values of $r_c$ and $\rho$ larger than those used in the main text, when the  clusterization condition $r_c\sqrt{\rho}>1$ is met. In contrast, for the largest interaction strength  and smallest temperature investigated in the main text glass behavior is not found by choosing $r_c=1$,  at which value self-assembled clusters are absent. These results underline the importance of cluster formation  for the emergence of glassiness in model (1).

 \subsubsection{Glass behavior for $r_c\sqrt{\rho}>$1: increasing $r_c$}
\begin{figure}[b]
\centerline{\includegraphics[width = 0.9\columnwidth]{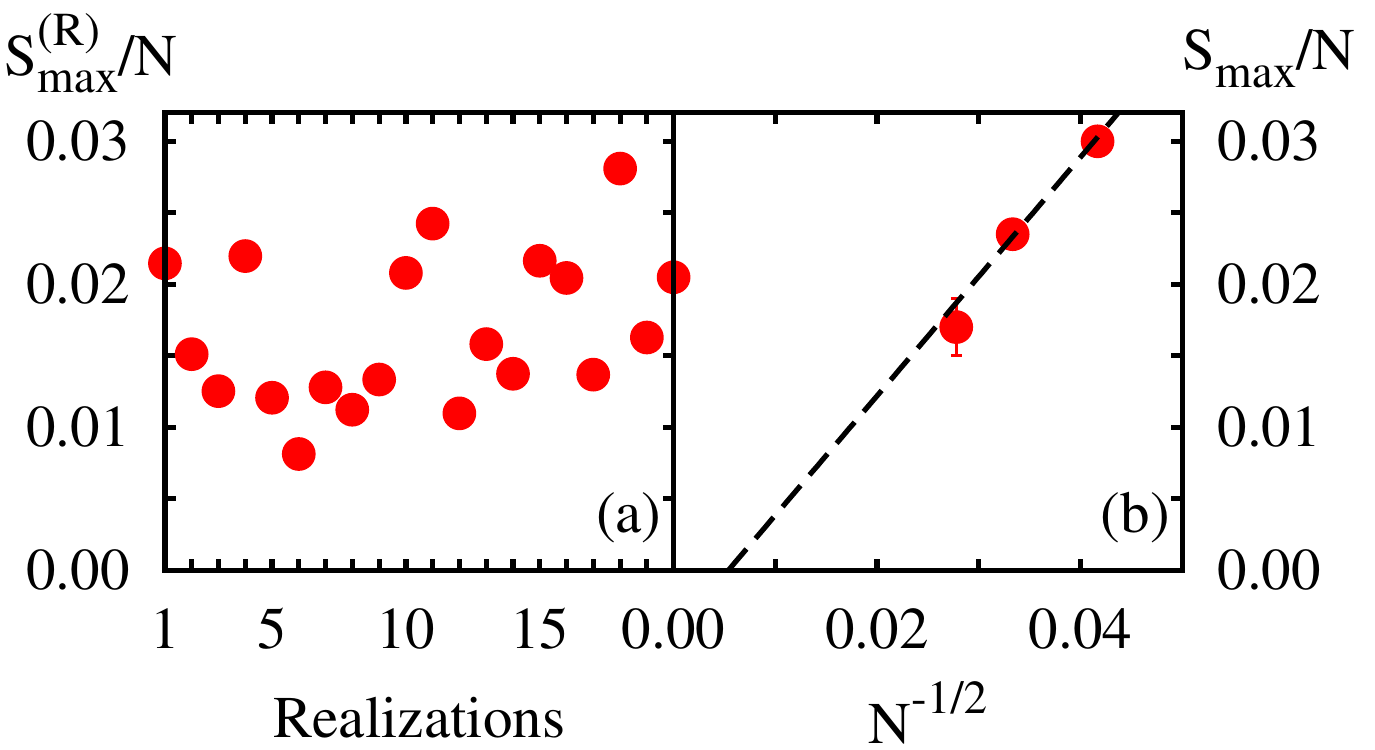}}
\caption{(color online). (a): Maximum value of the structure factor $S^{(R)}_{\max}/N$  for different realizations. The system size is $N=1296$. (b): size dependence of the average structure factor $S_{\max}/N$. In both panels $\rho=13/36$, $T/t=1/12$ and $V/t=10$. The hamiltonian of the system is that in Eq. (1) of the main text with $r_c=3$.} 
\label{FigS1}
\end{figure}
Figure~\ref{FigS1}(a) shows results for the maximum value of the structure factor $S^{(R)}_{\max}/N$  for different realizations of a low-temperature quench for interaction radius $r_c=3$ (larger than that in the main text) with particle density $\rho=13/36$ (same of that in the main text). The interaction strength  is chosen $V/t=10$, the temperature $T/t=1/12$, and the system size  $N=1296$. 

We find that, despite the relatively small value of $N$,   $S^{(R)}_{\max}/N$ clearly fluctuates between different realizations [Fig.~\ref{FigS1}(a)]. In addition, panel (b) shows that  the structure factor averaged over the individual realizations $S_{\max}/N$ vanishes  in the thermodinamic limit. In the same limit, the superfluid fraction is $\rho_s \simeq 0$, while the (realization averaged) Edwards-Anderson parameter $\tilde{q}_{EA}$ stays finite. These results demonstrate the existence of a glass phase with $r_c=3$. Here we have utilized the same parameters of the glass phase for  $r_c=2$ in the main text, showing that in this case an increase of $r_c$ does not alter the physical scenario.

 \subsubsection{Glass behavior for $r_c\sqrt{\rho}>$1: increasing $\rho$}
\begin{figure}[t]
\centerline{\includegraphics[width = 0.9\columnwidth]{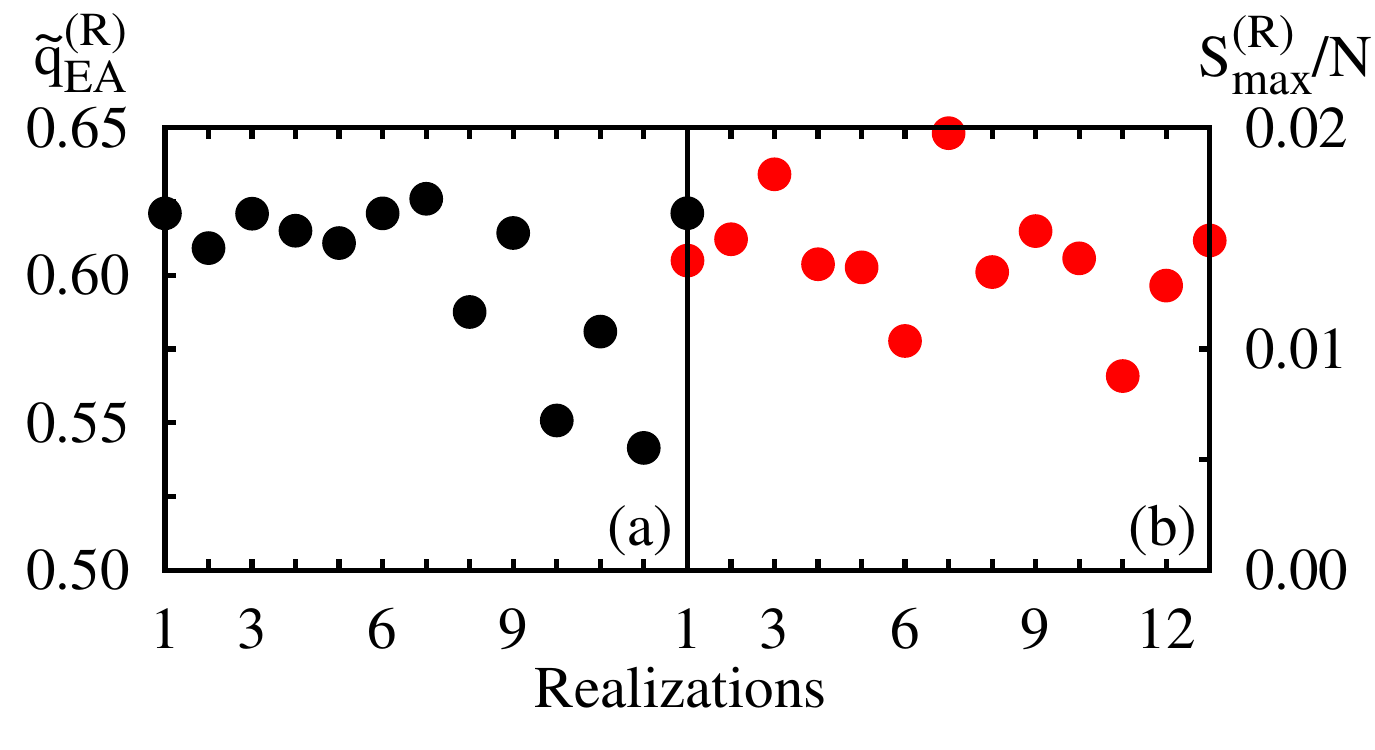}}
\caption{(color online). Edwards-Anderson parameter $\tilde{q}^{(R)}_{EA}$ (a)  and maximum value of the structure factor $S^{(R)}_{\max}/N$ (b)   for different realizations. The system size is $N=2304$, $\rho=0.3650$, $T/t=1/12$ and $V/t=5.4$. For these parameters the system is a superglass in the thermodynamic limit (see text).} 
\label{FigS2}
\end{figure}
Figure~\ref{FigS2} shows values of the Edwards-Anderson parameter $\tilde{q}^{(R)}_{EA}$ and $S^{(R)}_{\max}/N$ for different realizations of a temperature quench with target $T/t=1/12$, $V/t=5.4$,  $r_c=2$ (as in the main text) and $\rho=0.3650$. The latter  corresponds to a  variation at the percent level of  the mostly used density value in the main text (i.e., $\rho=13/36$). The system size is $N = 2304$.

Also in this parameter regime the maximum value of the structure factor [panel (b)] depends on the quench realization, showing the same glassy behaviour we observed at $\rho = 13/36$ [see main text, Figure 3, panels (b-c)]. The fluctuations of $\tilde{q}^{(R)}_{EA}$ are much less pronounced [panel (a)] and the values remains large. In addition, we find that here structural disorder and inhomogeneity coexist with finite superfluid fraction demonstrating the existence of a superglass. \\

At sufficiently large $V/t$, the existence of glassy phases for the cluster-forming potential in Eq. (1) of the main text has also been verified for values of the particle density fulfilling the clusterization condition $r_c\sqrt{\rho}> 1$ as high as $\rho \simeq 0.4$. For this value of $\rho$, an example of structural disorder and cluster formation is given in Fig. 4(d) of the main text. \\

Given the results above, as Hamiltonian (1) is particle-hole symmetric, one may expect that glass behavior could be observable for essentially all densities with  $r_c\sqrt{\rho}> 1$, $r_c=2$ and an appropriate choice of $V/t$. The investigation of this point and in particular of the scenarios emerging for very large values of $r_c\simeq \sqrt{N}/2$ (where the interaction range becomes of the order of the whole system size) will be the subject of a future work.

\subsubsection{Absence of glassy phases for $r_c = 1$ (no cluster formation)}
\begin{figure}[tb]
\centerline{\includegraphics[width = 0.9\columnwidth]{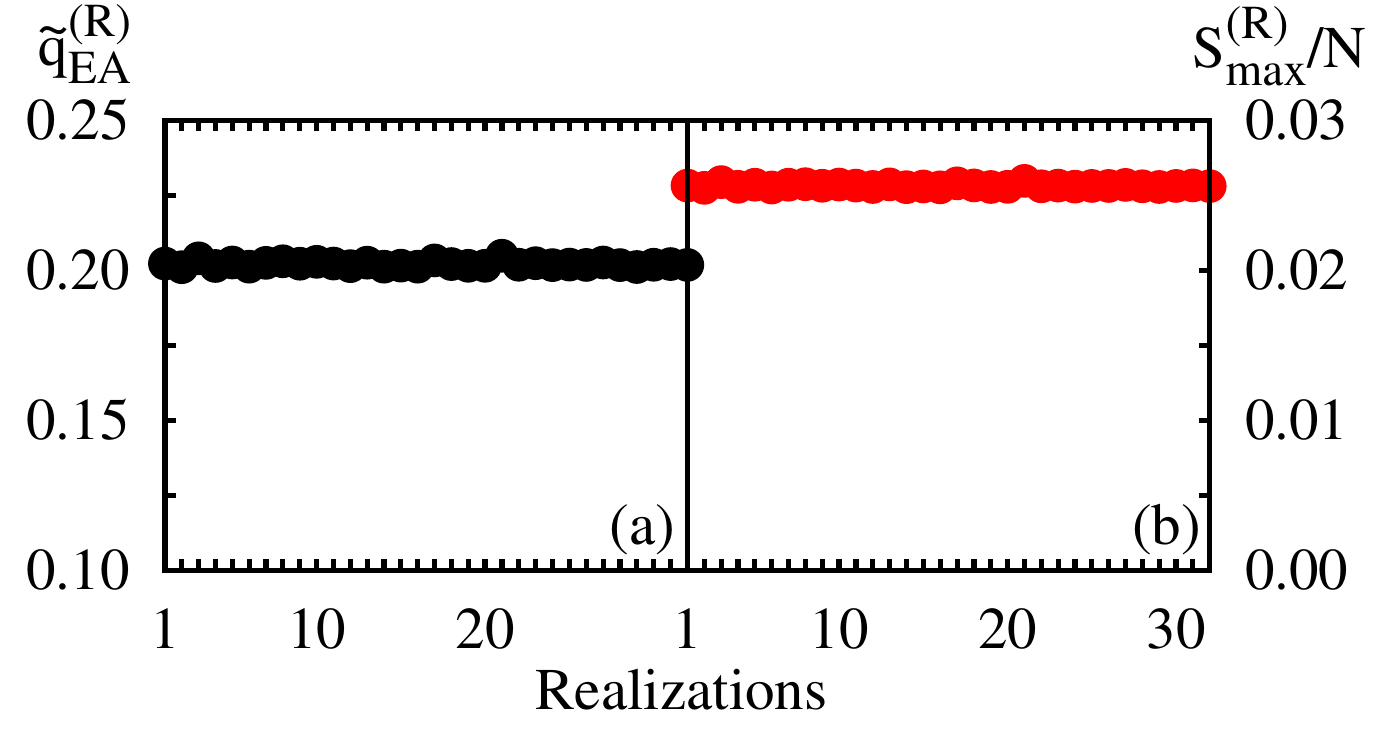}}
\caption{(color online). Same observables as in Fig.~\ref{FigS2} for $T/t=1/12$, $V/t=10$, $N=2304$ and $r_c=1$. In this case cluster formation does not occur. As opposed to the $r_c=2$ case where cluster formation induces disorder and result in an out-of-equilibrium superglass, here a temperature quench does not alter the equilibrium supersolid phase.}
\label{FigS3}
\end{figure}

Finally, to underline how cluster formation plays a crucial role for the observation of the glassy scenarios investigated here, we show in Fig.~\ref{FigS3} estimates of $\tilde{q}^{(R)}_{EA}$ and $S^{(R)}_{\max}/N$ for different realizations of a temperature quench with target $T/t=1/12$, $V/t=10$, $N=2304$ and $r_c=1$.
With this choice of parameters the interaction potential does not support clusterization and the equilibrium phase is a supersolid \cite{Boninsegni2005sm}. 

As shown in figure, the values of $\tilde{q}^{(R)}_{EA}$ and $S^{(R)}_{\max}/N$ are essentially identical in all realizations, i.e., the quench is ineffective and the equilibrium physics is restored. Conversely, at the same temperature, for $r_c=2$ and particle density satisfying the clusterization condition,  we find that the quenched counterpart of a supersolid is a superglass (see Fig.~2 main text). This indicates that cluster formation enhances frustration in the system favoring glassy behavior.

 \subsection{Glasses with soft-shoulder repulsive van-der-Waals interactions}
In general the phases studied in this work should be relevant for the whole large class of interaction potentials leading to cluster formation in the classical ground state \cite{Mladek2006sm}. 
Among these potentials, an interesting example is $V_{\rm ryd}(r)=V_{\rm ryd}/[1+(r/r_c)^6]$, which is realizable with cold Rydberg atoms~\cite{Gallaghersm,Molmersm,Pilletsm,PfauRevsm,RydbergExpsm}, where the ground state of each atom is off-resonantly coupled to an excited Rydberg state using a laser with (effective) Rabi frequency $\Omega$ and red detuning $\Delta$, with $\Delta\gg\Omega$~\cite{Santossm,Pupillo2010sm,Henkel2010sm,Honersm,Henkel2012sm,Alexsm,Macri2014sm,vanBijnensm}. In the appropriate parameter regime, $V_{{\rm ryd}}=\Omega^4/(16\pi\Delta^3)$ and $r_c=(C_6/2\Delta)^{1/6}$ is usually of the order of a few $\mu$m, with $C_6$ the coefficient of van-der-Waals type interactions for the atoms. We have checked numerically that   glassy phases similar to those described above can be obtained for $V_{\rm ryd}$ sufficiently large. As an example, Fig.~4(e) (main text) shows results for the density within the glass phase with $V_{\rm ryd}/t=30$. Below we give an example of possible parameters to achieve the regime of strong interactions in an optical lattice. While many more schemes for Rydberg dressing are currently being developed~\cite{Alexsm,vanBijnensm,Whitlock2015sm}, we notice that first examples of Rydberg-dressing have been very recently realized in three different experiments~\cite{Biedermann2015sm,Killian2015sm,Grosssm}. \\

Here we consider a gas of $^{85}$Rb atoms, where the groundstate 
$|g\rangle$ of each atom is coupled off-resonantly to an excited Rydberg
state $|r\rangle$ using a (two-photon) laser with (effective) Rabi frequency $\Omega$
 and detuning $\Delta$ from the atomic transition. We choose red detuning in order to realise the wanted soft-shoulder potential~\cite{Henkel2010sm,Cinti2010sm} and
 $|\Delta |\gg \Omega$, in order to satisfy the weak dressing condition. As an example, we consider $|r \rangle \equiv |43 S_{1/2}\rangle$,
 $\Omega / (2\pi)= 2.0$MHz and $\Delta/(2\pi) = 20$ MHz, where we
have set Planck' s constant $\hbar=1$. These
parameters imply $V_{\rm ryd}\simeq 450$ Hz. 
 The resulting cut-off radius $r_c > 2 a$, with $a \simeq 0.5 \mu$m the characteristic length for the lattice spacing, allows one to easily access the cluster regime. The hopping amplitude $t$ depends exponentially on the depth of the optical lattice~\cite{Zwergersm},
 which allows one to achieve the regime of strong interactions $V/t$. For example, for an in-plane lattice depth $V_0/E_R \gtrsim13$, $E_R=2\pi\times$ 50 kHz, $t/(2\pi)\lesssim 40$ Hz, and thus $V/t \gtrsim 10$. This value can be further increased by increasing $V_0$. A depth $V_\perp$ of the optical lattice in the transverse direction with $V_\perp \gg V_0$ ensures that the dynamics is two-dimensional.
 Here, $E_R/(2\pi)=\pi^2/(2ma^2)$ is the recoil
energy of atoms with mass $m$. 
Finally, the small effective decay rate $\gamma_{\rm eff}/(2\pi) = (\Omega/\Delta)^2\gamma_r$ of the dressed ground-state results in a
$\gamma_{\rm eff}$ usually of the order of tens of Hz, with
$\gamma_r$ the bare decay rate of the Rydberg state. While this small rate can be still detrimental for reaching 
equilibrium phases, as noted in the main text the glassy phases of interest here are obtained out of equilibrium. The ensuing  dynamics in the glassy phase after the quench is essentially frozen and could be observed immediately after the quench. We further note that, as shown above, variations of the particle number by a few percents does not hinder the realization of glassy phases.Provided that residual spontaneous emission from the Rydberg states results only in single-particle losses~\cite{textnote}, we  expect that glassy phases should be observable on a time scale sufficiently larger than $t$ and 1$/\gamma_{\rm eff}$.

These aspects should considerably simplify the  detection of glassy phases in experiments. 
 \subsection{Quench in $V/t$ at fixed $T$}

 Since in experiments it should be easier to perform a quench in $V/t$ (which, e.g., with Rydberg atoms would entail
modifying the laser parameters for Rydberg dressing, changing $V$, or for the depth of the confinement to the optical lattice, changing $t$) than in $T$, we have investigated
 whether the glassy behaviour described above is also found upon a quench in the interaction strength.\\
 
\begin{figure}[t]
\centerline{\includegraphics[width = 0.9\columnwidth]{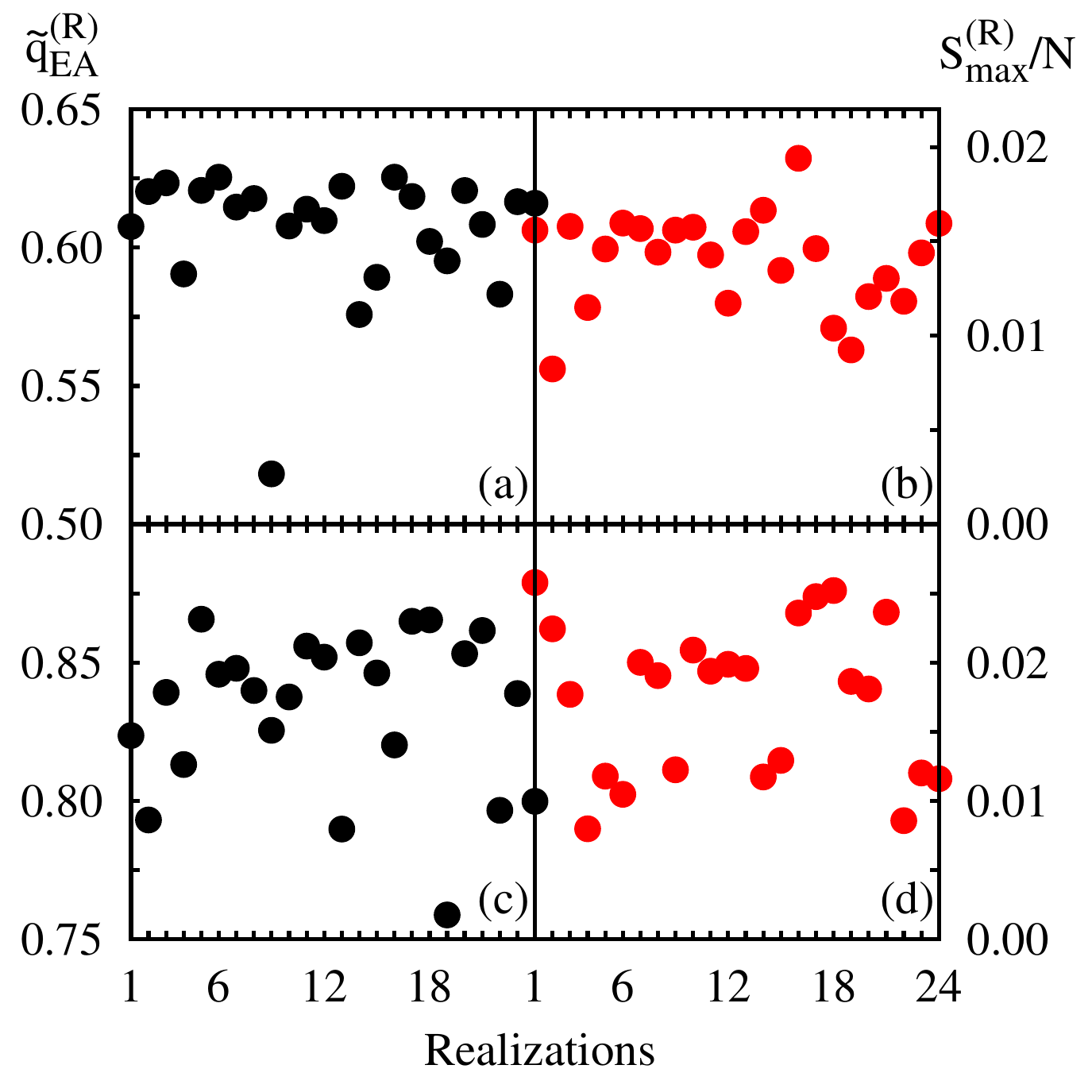}}
\caption{(color online). Same observables as in Fig.~\ref{FigS2} when the simulation protocol is based on a quench in the interaction strength (see text). Here $\rho=13/36$, $N = 2304$ and $T = 1/12$. The final value of the interaction strength is $V/t=5.4$ [panels (a) and (b)] and $V/t=10$ [panels (c) and (d)]. In the thermodynamic limit the former  choice of $V/t$ leads to a superglass, the latter, to a normal glass.}
\label{FigS4}
\end{figure}

We find that similar behavior of $\tilde{q}^{(R)}_{EA}$ and $S^{(R)}_{\max}/N$ to the one above is obtained when, starting from a value at which the system is superfluid, $V/t$ is abruptly increased to an intermediate or large value (see Fig.~\ref{FigS4}). Following this simulation protocol at fixed (low enough) $T/t$ (i.e., a quench in $V/t$)  we  find no crystalline order and inhomogeneity in the thermodynamic limit. Specifically, the system is a superglass or  a normal glass for intermediate or large values of $V/t$, respectively.  This is entirely analogous to what found in the main text and above for a quench in $T$.


\begin{thebibliography}{}
\bibitem{textnote} We note that collective decay mechanisms \cite{Grosssm} linked to the population of excited Rydberg states \cite{Porto} have been very recently demonstrated. In principle, these effects may be mitigated by applying additional lasers, that efficiently depopulate these states, on relevant transitions.
\end{thebibliography}
\end{document}